\documentclass[doublecol]{epl2}

\begin{document}

\title{Dynamical thermalization of Bose-Einstein condensate\\
 in Bunimovich stadium}

\author{L.Ermann\inst{1}, E.Vergini\inst{1} \and  D.L.Shepelyansky\inst{2}} 
\shortauthor{L.Ermann, E.Vergini, D.L.Shepelyansky}
\institute{
 \inst{1} Departamento de F\'{\i}sica, Gerencia de Investigaci\'on y Aplicaciones,
 Comisi\'on Nacional de Energ\'{\i}a At\'omica.
 Av.~del Libertador 8250, 1429 Buenos Aires, Argentina\\
 \inst{2} Laboratoire de Physique Th\'eorique du CNRS (IRSAMC), 
  Universit\'e de Toulouse, UPS, F-31062 Toulouse, France
}
\date{May 18, 2015}

\abstract{We study numerically the wavefunction evolution 
of  a Bose-Einstein condensate
in a Bunimovich stadium billiard being governed by the Gross-Pitaevskii
equation. We show that
for a moderate nonlinearity, above a certain threshold,
there is emergence of dynamical thermalization which
leads to the Bose-Einstein probability distribution
over the linear eigenmodes of the stadium. 
This distribution is drastically different from the energy
equipartition over oscillator degrees of freedom which would 
lead to the ultra-violet catastrophe.
We argue that this interesting phenomenon can be studied 
in cold atom experiments.
}
%


\maketitle

\section{Introduction}

Starting from the famous Fermi-Pasta-Ulam problem \cite{fpu1,fpu2}
the interest to understanding of thermalization process
in dynamical systems with a finite number of degrees of freedom
is continuously growing (see e.g. \cite{fpu3,fpu4}).
At present the experiments with cold Bose gas
in atom traps and optical lattices 
open access to experimental investigations
(see e.g. \cite{bloch,langen}) stimulating the theoretical
and experimental interest to this  phenomenon
(see e.g. \cite{castin1,castin2}).

The numerical analysis of dynamical thermalization 
in disordered nonlinear chains
has been started recently showing that 
the quantum Gibbs distribution can appear at a moderate nonlinearity
contrary to usually expected energy equipartition over linear
modes \cite{mulansky,njpermann}. Thus it is interesting to analyze 
the dynamics of a Bose-Einstein condensate (BEC),
described by the Gross-Pitaevskii Equation (GPE),
in a chaotic billiard where 
the quantum evolution corresponds to a regime of quantum chaos and
energy levels statistics described 
by the Random Matrix Theory \cite{bohigas,stockmann,haake}. 
As an example we use a de-symmetrized
Bunimovich stadium where the classical dynamics
is fully chaotic (see e.g. \cite{bunimovich,stockmann}).  
We note that the chaotic optical billiards, created by a laser beam
and containing cold atoms,
have been already studied experimentally
\cite{raizen2001,davidson2001} and hence our 
model can be investigated experimentally.

\section{Model description}
\begin{figure}[h]
\begin{center}
\includegraphics*[width=7.8cm]{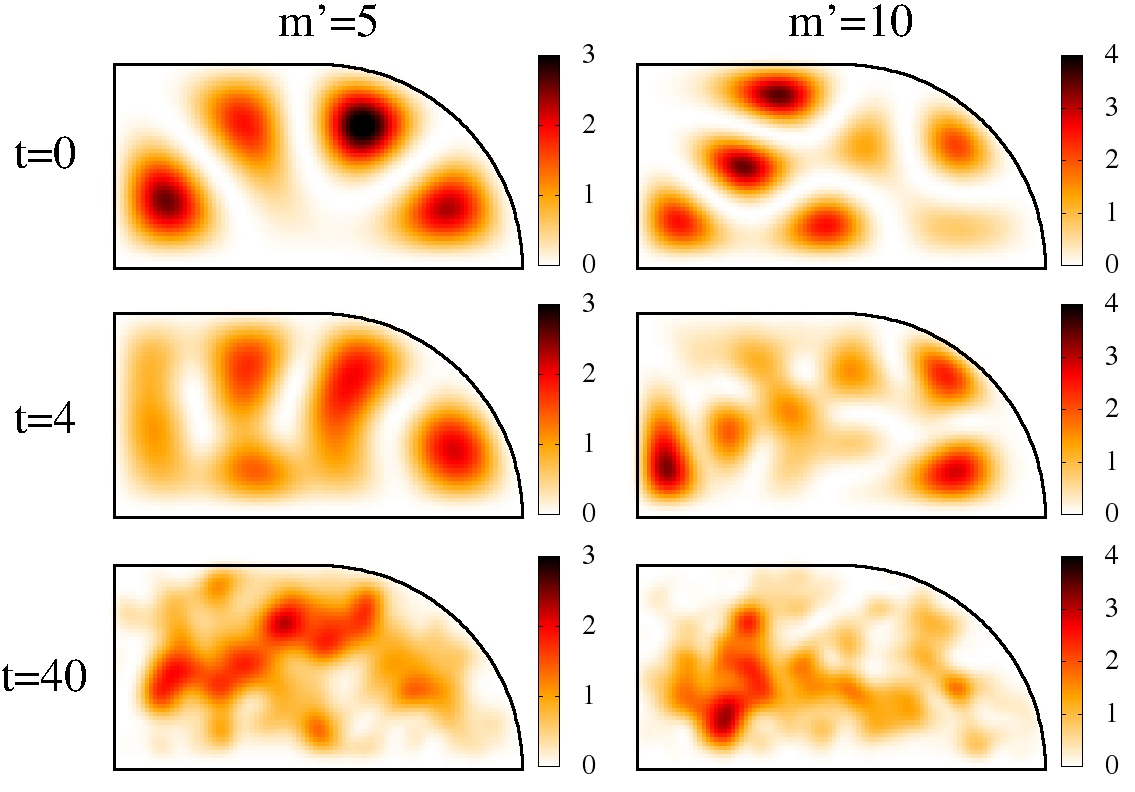}
\end{center}
\vglue -0.3cm
\caption{(Color on-line) 
Time evolution of 
wavefunction amplitude 
 $\vert\psi(x,y,t)\vert$
in coordinate 
representation for an initial state 
of linear eigenstate mode $m'=5$ (left column)
and  $m'=10$ (right column)  shown at $t=0$ (top panels).
Middle and bottom panels show the snapshots of
corresponding distributions at 
$t=4$ and $t=40$ respectively. Here $\beta=10$, color bars give
values of $\vert\psi(x,y,t)\vert$. }  
\label{fig1}
\end{figure}

The model is described by the GPE 
for BEC in the  de-symmetrized Bunimovich stadium billiard 
with Dirichlet boundary conditions:
\begin{equation}
\label{eq1}
 i\hbar{\partial\psi(\vec{r},t)\over\partial t}=-{\hbar^2\over 2m}
 \Delta \psi(\vec{r},t)+\beta \vert\psi(\vec{r},t)\vert^2\psi(\vec{r},t)
\end{equation}
where we consider $\hbar=1$, $m=0.5$. The height of the stadium is taken as $h=1$ and its
maximal length is $l=2$ (see Fig.~\ref{fig1}). Thus the average level spacing is 
$\Delta \approx 4\pi/A \approx 7.04$ where $A$ is the billiard area. 
At $\beta=0$ the numerical methods of quantum chaos allows to
determine efficiently about million of eigenenergies of linear modes
and related eigenmodes \cite{vergini}.
For comparison, we also consider  the case on a rectangular billiard 
with approximately the same area as for stadium
and with the  golden mean ratio $l/h=(1+\sqrt{5})/2, h=1$.
We note that for model (\ref{eq1}) the spectrum of Bogoliubov excitations of BEC
in a Bunimovich stadium had been studied in \cite{raizen2004},
but the question of dynamical thermalization has been not addressed there.
%
We also point that the model (\ref{eq1}) is described by the partial 
differential equation (continuous variables)
thus being significantly more complex than 
the case of nonlinear chains studied in 
\cite{mulansky,njpermann}. Indeed, even the prove of the existence of  
solution in (\ref{eq1}) is a nontrivial mathematical problem
which still remains open
(e.g. the ultra-violet catastrophe would imply the absence of solution).
In the following we restrict our analysis to the GPE case not
going beyond this description.


The time evolution of (\ref{eq1}) is integrated
by small time steps of Trotter decomposition of linear and nonlinear terms
with a step size going down to  $\Delta t =4 \times 10^{-5}$ to suppress nonlinear 
growth of high modes. We use $N_c=1076 (1085)$ linear 
modes $\phi_m$ of stadium (rectangular)
for linear part of time propagation doing the nonlinear step with $\beta$ term
in the coordinate space with $N_p=11207 (12816)$  points inside the billiard.
The lattice points are given by $79\times79=6241$ equidistant $x$-$y$ coordinates 
for the square part of the stadium billiard 
(rectangular billiard), and rays with equidistance in angles for the circular part.
The change of basis from coordinates to energies (and viceversa) is given by a unitary matrix
in double precision.
Similar to  \cite{kolmoturb}, a special aliasing procedure
is used with an efficient suppression of nonlinear numerical instability at high modes. 
This integration scheme exactly conserves the probability norm
providing the total energy conservation with an accuracy better than $2\%$
at largest value $\beta=20$ and better than $1\%$ at lower $\beta$ values.
At any step the wavefunction is expanded in the basis of linear
modes $\phi_m$ so that $\psi(x,y,t)= \sum_m C_m(t) \phi_m(x,y)$.
The averaging of probabilities $w_m(t)=\vert C_m(t) \vert^2$ ($\sum_m w_m=1$)
over time gives the average probability distribution 
$\rho_m = \langle\vert C_m\vert^2\rangle_t$.

We note that the quantum evolution of GPE (\ref{eq1})
has been studied in the frame of quantum turbulence
for a rectangular billiard
\cite{nazarenko} and for a 3D-cube  \cite{tsubota}. 
However, in these studies there is energy injection/absorption at
low/high modes to generate the Kolmogorov energy flow in space of 
linear modes (see e.g. \cite{zakharov,nazarenkobook}). 
We also note that 
the time evolution of 
wave packet for the GPE in Bunimovich stadium had
been simulated in \cite{xiong2010}
but only a spacial distribution had been considered there.
In contrast to these studies we consider
only unitary or Hamiltonian evolution given by (\ref{eq1})
being interested in the distribution properties of 
probabilities $\rho_m$ over linear eigenmodes. 
In this respect our approach is different from other studies
where the analysis had been concentrated on space fluctuations
(see e.g. \cite{xiong2011}).
Also, as we will see below, there is
a significant difference for the GPE evolution 
in chaotic and rectangular billiards.

\section{Time evolution}
Examples of time evolution for two initial eigenmodes 
$m'=5, 10$ are shown in Fig.~\ref{fig1} 
at $\beta=10$ (video is available at \cite{ourwebpage}).
They show that, due to nonlinearity
inside the stadium, there are complex 
irregular oscillations of wavefunction with time.

\begin{figure}[h]
  \includegraphics[width=0.47\textwidth]{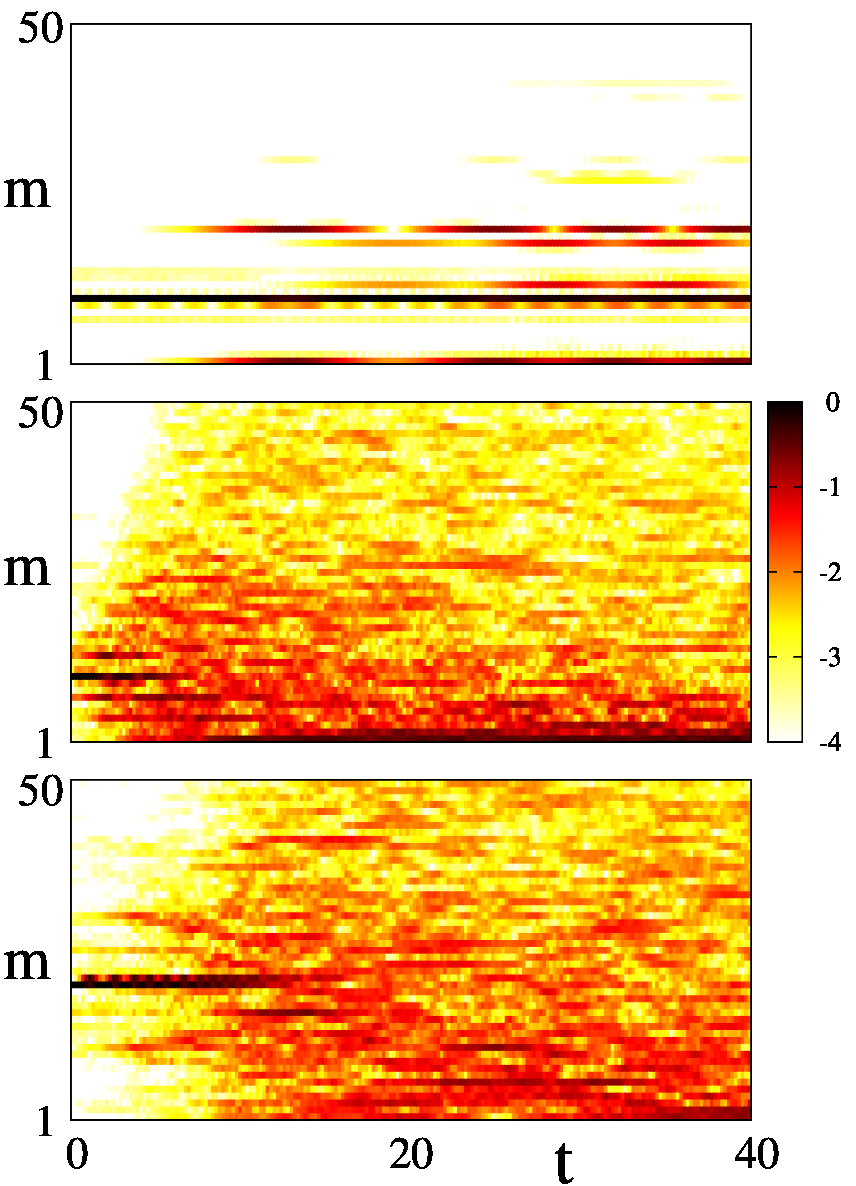}
 \caption{(Color on-line)
Time evolution of probabilities
$w_m(t)$  in linear mode basis 
for initial state $m^\prime=10$ at $\beta=2$ (top panel),
$m^\prime=10,20$ at $\beta=10$ 
 (middle and bottom panels respectively).
The probabilities $w_m(t)$ are averaged over time interval
$\delta t =0.2$ to reduce fluctuations; color bar shows 
values of $\log_{10} \rho_m(m')$.
 }
 \label{fig2}
\end{figure}

Another representation is obtained by
considering the time evolution 
of probabilities $w_m(t)$ in the basis of linear modes shown in Fig.~\ref{fig2}. 
At moderate value $\beta =2$ the probability remains located only in a few 
modes without thermalization and spreading over many modes.
For larger value of $\beta =10$ the nonlinear
spearing over modes goes in a more efficient way with many
excited modes. Thus the dynamical thermalization is expected
to be absent at small or moderate $\beta < \beta_c \sim 1$
while for large nonlinearities $\beta \sim 10 > \beta_c$
we may expect the emergence of dynamical thermalization. 

For the initial state $m'=10$ we have the approximate energy value
$E_m \approx m' \Delta \approx mv^2/2 \approx 70$, 
where $v \approx \sqrt{280}$ is a velocity
of classical particle and a time interval between
collisions is approximately $\tau_{col} \approx h/v \approx 0.06$. 
Thus during the time $t =40$ we have approximately 
$N_{col} = t/\tau_{col} \approx 670$.
Dynamical thermalization is reasonably achieved for time 
interval $t\in[20,40]$ as it is visible in middle and bottom panels 
of Fig.\ref{fig2} where $w_m(t)$ have thermalized like distribution 
with $\beta=10$.
%

\section{Bose-Einstein thermal distribution}

To characterize the dynamical thermalization in more detail
we assume that a moderate nonlinearity acts as a certain 
thermalizer which drives the system to a thermal equilibrium
over quantum levels of the stadium. At the same time we assume
that the nonlinear term is not very strong
so that it does not affect significantly the average
linear eigenenergies. Indeed, on average we have
$\beta \vert \psi \vert^2 \approx \beta/A  \approx \Delta$
for $\beta \sim 10$, so that indeed, the nonlinear energy shift is moderate
at such values of $\beta$.

\begin{figure}[h]
  \includegraphics[width=0.47\textwidth]{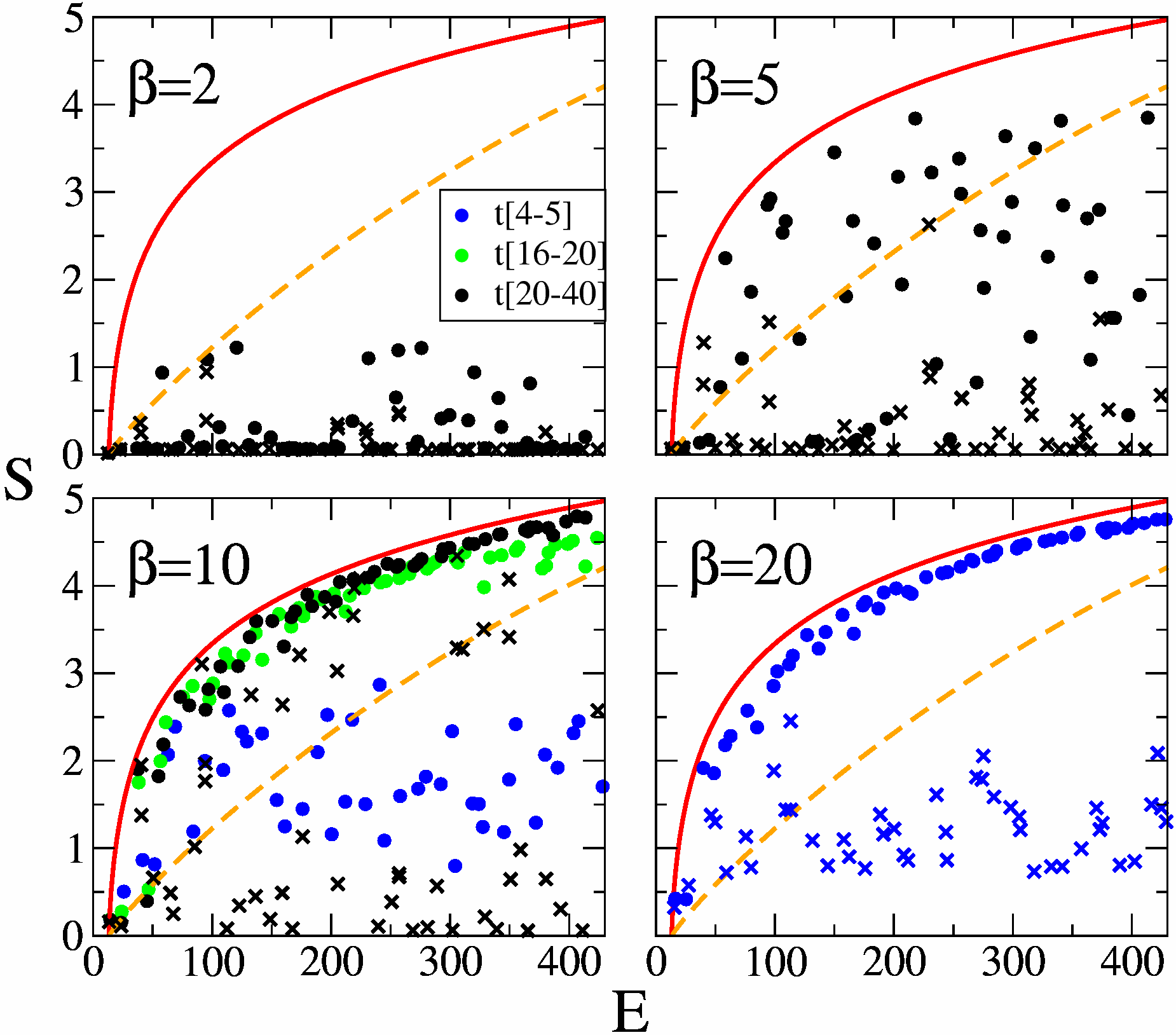}
 \caption{(Color on-line) 
Entropy dependence on energy $S(E)$ 
obtained from the GPE time evolution of
initial linear eigenstates with
$1 \leq m' \leq 50$ for the stadium (circles) 
and rectangular (crosses) billiards
for nonlinearity $\beta=2, 5, 10, 20$
marked on each panel. Here
the  average is done over time intervals
$t\in[4,5]$ (blue), $t\in[16,20]$ (green) and 
$t\in[20,40]$ (black). The red curve
represents the Bose-Einstein ansatz (\ref{eq2})
 while the orange dashed 
curve shows the case of 
energy equipartition over first 50 modes of 
the stadium.
 }
 \label{fig3}
\end{figure}

\begin{figure}[h]
  \includegraphics[width=0.47\textwidth]{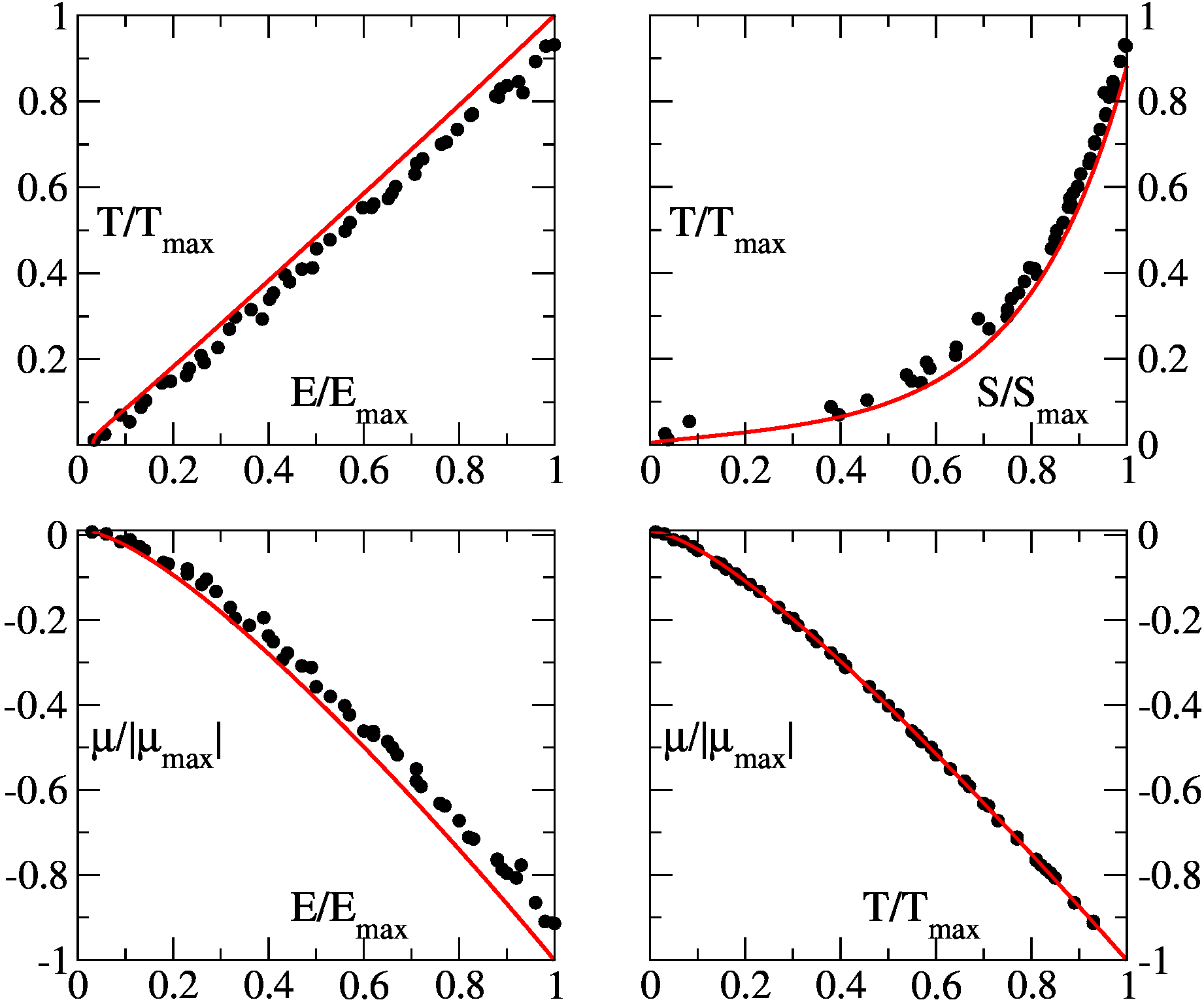}
 \caption{(Color on-line) 
Theoretical dependencies given  by
Bose-Einstein ansatz (\ref{eq2}) and shown by
the red curves
for $T(E)$ (top left panel),
$T(S)$ (top right panel),
$\mu(E)$ (bottom left panel),
$\mu(T)$ (bottom right panel);
the numerical results, obtained from 
GPE in the stadium, are shown by 
black points at $\beta=10$
with averaging over the time interval $[20,40]$.
For representation convenience 
we show these dependencies using
rescaling to maximal values of variables
corresponding to initial state with
$m'=50$: $E_{max}=414$, $S_{max}=4.8$, 
$T_{max}=387.75$, $\vert\mu_{max}\vert=1500.45$.
}
 \label{fig4}
\end{figure}

Thus we expect that nonlinearity generates a dynamical thermalization over
the quantum billiard energy levels. In such a case we should have the standard
Bose-Einstein distribution ansatz  over energy levels $E_m$ \cite{landau}:
\begin{equation}
\label{eq2}
 \rho_m=1/[\exp[(E_m-E_g-\mu)/T]-1]
\end{equation}
where $E_g = 13.25$ is the energy of the ground state,
$T$ is the temperature of the system,
$\mu(T)$ is the chemical potential dependent on temperature.
The values $E_m$ are the eigenenergies of the stadium at $\beta=0$.
The parameters $T$ and $\mu$ are determined
by the norm conservation
$\sum_{m=1}^{\infty} \rho_m =1$ (we have only one particle in the system)
and the initial energy $\sum_m E_m \rho_m =E$.
The entropy $S$ of the system is determined by
the usual relation \cite{landau}: $S= - \sum_m \rho_m \ln \rho_m$.
The relation (\ref{eq2}) with normalization condition 
determines the implicit dependencies on temperature
$E(T)$, $S(T)$, $\mu(T)$.

The advantage of  energy $E$  and entropy $S$ 
is that both are extensive variables,
thus they are self-averaging and due to that 
they have reduced fluctuations.
Due to this feature $S$ and $E$ are especially convenient for
verification of the thermalization  ansatz. To check this
ansatz we start from an initial linear mode $m'$ which
corresponds to the system energy $E \approx E_{m'}$ and follow the GPE time evolution
of probabilities $w_m(t)$ determining the value of entropy $S$ from obtained
average probabilities $\rho_m$. Considering the initial states with $1 \leq m' \leq 50$
we obtain the numerical dependence $S(E)$ shown by symbols in Fig.~\ref{fig3}.
This dependence is compared with the analytic curve
following from the Bose-Einstein ansatz (\ref{eq2})
which gives the dependencies $E(T)$ and $S(T)$ and hence provides
the analytic dependence $S(E)$ shown by the red curve
in Fig.~\ref{fig3}.

The data of Fig.~\ref{fig3} show that even at large $\beta=20$
there  is no thermalization for the rectangular billiard.
 We attribute this to the fact that the ray
dynamics is integrable in this billiard and thus
it is much more difficult to reach onset of 
chaos for the GPE in this billiard at moderate nonlinearity
studied here.

The situation is different for the stadium: at $\beta=2$ 
only a few modes are populated, at $\beta=5$ the number of modes
is increased but still the numerical data for
$S$ are very far from the thermalization red curve
(at least on the time scale reached in our numerical simulations).
 However, for $\beta =10, 20$ we find that the numerical data
at large times $t>15$ follow the theoretical 
curve $S(E)$ given by the Bose-Einstein thermalization.
A small visible deviation from theory
is still visible since the numerical points are systematically slightly
below the theory curve. We attribute this to a finite
computation time which apparently is not long enough
to visit all regions of multi-configurational space
with sufficiently large statistics. 
On the basis of obtained data
we can conclude that the dynamical thermalization
in the stadium sets in for $\beta > \beta_c \approx 7 \approx \Delta$.
We also checked that the initial states,
which represent a linear combination of a few
eigenmodes, also follow the theoretical red curve in Fig.~\ref{fig3}
at $\beta > \beta_c $ (e.g. two modes $m=10, 15$ at $\beta=10$).

Another way to check the thermalization predictions
is to determine $T$ from the numerical values of $E$, $S$
which, according to the Bose-Einstein ansatz, give 
independent values $T_1(E)$
and $T_2(S)$. The average value $T=(T_1+T_2)/2$
is shown in Fig.~\ref{fig4}
as a function of numerical values of $E$ and $S$. 
We see that the numerical points
are in a good agreement with the analytic curves
(apart of small
systematic displacement of numerical points discussed 
in the paragraph above). In a similar way
we make a comparison between the theory and numerical data
for dependencies $\mu(E)$ and $\mu(T)$ shown in bottom panels
of Fig.~\ref{fig3}. Again we find a good agreement
between the numerical data and the Bose-Einstein
thermalization distribution (\ref{eq2}).

\begin{figure}[h]
  \includegraphics[width=0.47\textwidth]{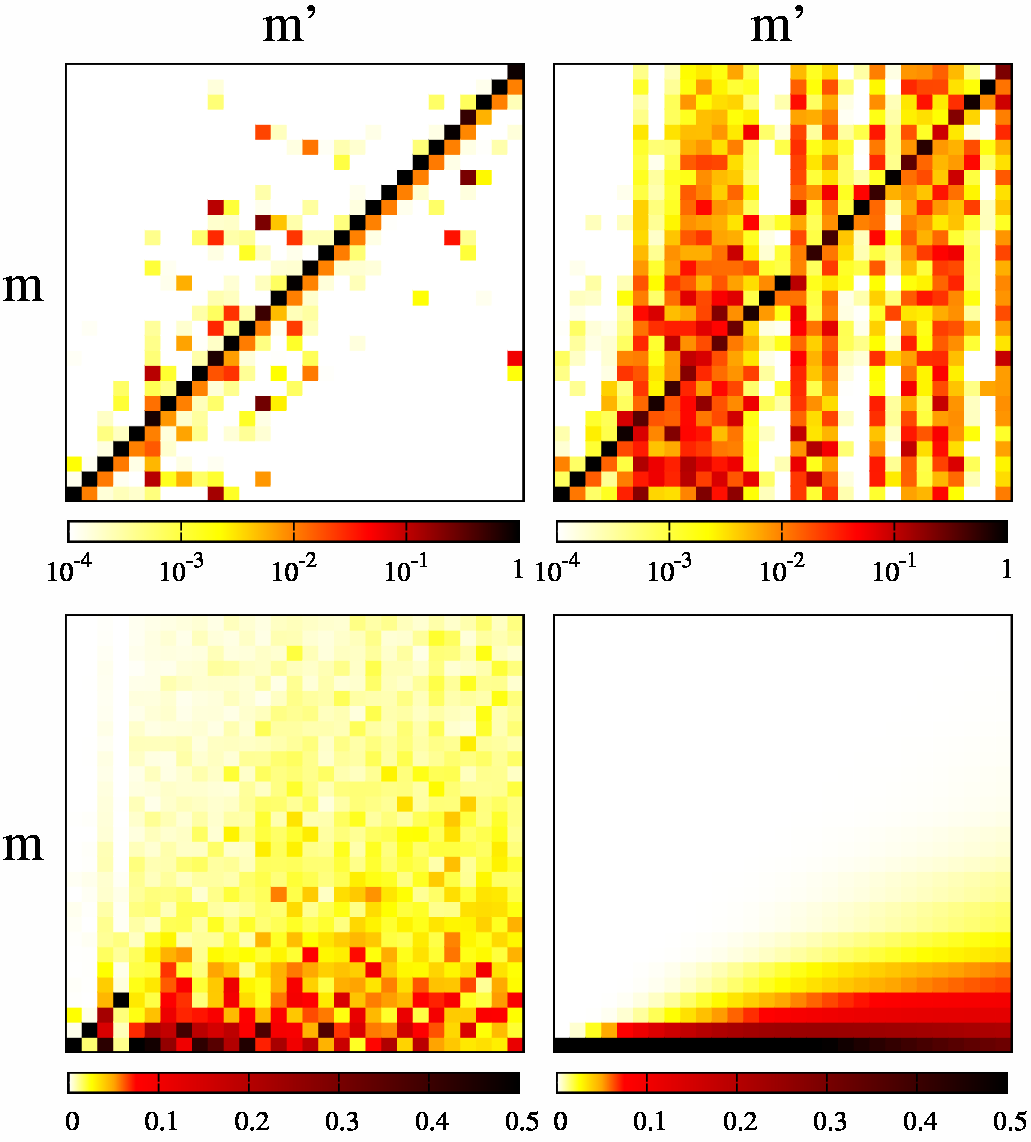}
 \caption{(Color on-line) Time averaged probabilities  
 $\rho_m(m^\prime)$ at stadium eigenstate $m$ for initial state $m^\prime$;
the time averaging is done for time intervals $[20,40]$;
the panels show data for $1 \leq m^\prime, m \leq 30$
in $x,y$ axes respectively. Here we show
the cases: $\beta=2$ (top left panel),
$\beta=5$ (top right panel),
$\beta=10$ (bottom left panel),
the theoretical Bose-Einstein distribution (\ref{eq2})
(bottom right panel).
The values of $\rho_m(m^\prime)$ 
are shown by color with the corresponding color bars for each panel.
 }
\label{fig5}
\end{figure}

The validity of the Bose-Einstein thermalization ansatz (\ref{eq2})
leads to a striking paradox pointed already for nonlinear chains in \cite{njpermann}:
formally the GPE in stadium 
gives a system of equations for nonlinear coupled oscillators
(we have nonlinear coupling between linear oscillator modes of the stadium)
with a moderate nonlinearity. The usual
expectations of the statistical mechanics predict the energy equipartition
between these modes \cite{landau,njpermann}. If the number of modes
is infinite then we should have ultra-violet catastrophe
with probabilities $\rho_m \sim 1/E_m$ at high modes $m$
and the global temperature approaching zero as $T \sim 1/m_{max}$
for initial excitation with $m^\prime \sim 10$
(here $m_{max}$ is the maximal mode number).
This classical thermalization ansatz for 
$m_{max}=50$ is shown by a dashed curve in Fig.~\ref{fig3}
and the data clearly show that it is very different
from the numerical data which are close to the Bose-Einstein
ansatz (we note that numerically the quantum Gibbs distribution \cite{njpermann} over 
quantum levels of stadium gives the results being 
rather close to those of (\ref{eq2}) since at low temperatures
and large $m$ values both distributions are rather similar).
Thus our data clearly show the emergence of the dynamical thermalization
described by the Bose-Einstein distribution in a chaotic billiard
for moderate nonlinearity $\beta > \beta_c \sim \Delta$.

A more detailed check of the Bose-Einstein distribution
requires a direct comparison of numerically obtained
probabilities $\rho_m(m^\prime)$ with the theoretical 
expression (\ref{eq2}) for each initially excited
mode $m^\prime$. We show such data in  Fig.~\ref{fig5}
for $\beta=2, 5, 10$. It is clear that there is no thermalization
at $\beta =2, 5$ since a large fraction of probability
remains at the initially populated state $m=m^\prime$.
For $\beta = 10$ we see that the probability at initial state
$m^\prime$ drops significantly indicating emergence 
of dynamical thermalization.
However, still the numerical probabilities at large $m$
have larger values compared to those of the theory (\ref{eq2})
shown in the bottom right panel of Fig.~\ref{fig5}.
We attribute this to the fact that our total computation time
is not large enough to have good statistical data for 
average values of $\rho_m(m^\prime)$ which require good
averaging and long computation times. Such a problem
had been visible in the numerical simulations
with nonlinear chains \cite{mulansky,njpermann}
where the time of simulations have been by a few orders 
of magnitude larger than here. At the same time the extensive property
of energy $E$ and entropy $S$ makes them 
self-averaging and more stable in respect to fluctuations
thus allowing to compare them with the theory (\ref{eq2})
at significantly shorter time scales.

Unfortunately the large scale simulations of the GPE for stadium
are rather heavy and time consuming since they require
transformations from coordinate to linear mode space
and small integration time step with aliasing procedure
to suppress numerical instabilities of high modes.
It is possible that the numerical codes can be improved 
allowing to reach larger time scales but 
this requires further studies going beyond the scope of this work.

Finally we discuss a preparation of one or a few  initial eigenstates
considered above for the time evolution and dynamical thermalization.
It is clear that the ground state of the billiard is
relatively easy to prepare since it is the final state
in a process of relaxation and also since it is 
compact in space being close to a coherent state of
a harmonic trap. An excited state can be produced from the
ground state applying a monochromatic driving (oscillation)
of the billiard that creates an effective
{\it ac}-potential $V_{ac}= f x \cos(\omega t)$ 
if one goes to the oscillating frame
(see e.g.   \cite{prosen}). In a chaotic billiard dipole
matrix elements have transitions between all  energy eigenstates 
\cite{prosen} and thus a resonant transition will populate
one or a few states being close to the resonance
$E_n \approx E_0 + \hbar \omega$. We note that such a method
demonstrated already its efficiency for excitation
of high energy states for chaotic Rydberg atoms 
(see e.g. \cite{delande,koch}).

\section{Discussion}

Our studies of the GPE in the Bunimovich stadium billiard
show that for a moderate nonlineariy parameter above a certain threshold
$\beta > \beta_c \approx \Delta$ the nonlinear Hamiltonian
dynamics leads to emergence of dynamical thermalization
over the linear billiard modes which is well described 
by the Bose-Einstein distribution. This distribution
is strikingly different from the usually expected
energy equipartition over modes \cite{landau,njpermann}
which would lead to a violet catastrophe
with a significant probability transfer to
higher and higher modes of the chaotic billiard.
The established validity of the  Bose-Einstein distribution,
together with the previous studies of dynamical thermalization
in nonlinear chains \cite{mulansky,njpermann},
leads to an unexpected conclusion 
about emergence of quantum distributions
over linear energy modes in 
systems of coupled nonlinear oscillators at moderate nonlinearity.
This result is drastically different from the standard
energy equipartition picture expected for
nonlinear dynamics of oscillator systems 
\cite{fpu1,fpu2,fpu3,fpu4,landau}.

The described picture of ``quantum'' dynamical thermalization
for the GPE in chaotic billiards requires a better understanding of
nonlinear dynamics in systems with many degrees of freedom.
It is known that slow chaos, like the Arnold diffusion 
\cite{chirikov1979,chirikov1997,mulansky2},
and an anomalous diffusion in disordered nonlinear chains
(see e.g. \cite{kolmoturb,danse,fishman,flach}) generate a number of features
which still wait their deep understanding.
We hope that our results will stimulate further research in this
field of fundamental aspects of nonlinear dynamics and thermalization
onset in systems with large but finite number of degrees of freedom.

The modern progress in the cold atom experiments
allows to investigate a dynamics of Bose gas and Bose-Einstein condensates
\cite{bloch,langen} while the chaotic billiard for such
atoms can be created by optical beams \cite{raizen2001,davidson2001}.
Thus we think that the model (\ref{eq1}) can be realized with cold
atom experiments.

\begin{figure}[h]
  \includegraphics[width=0.47\textwidth]{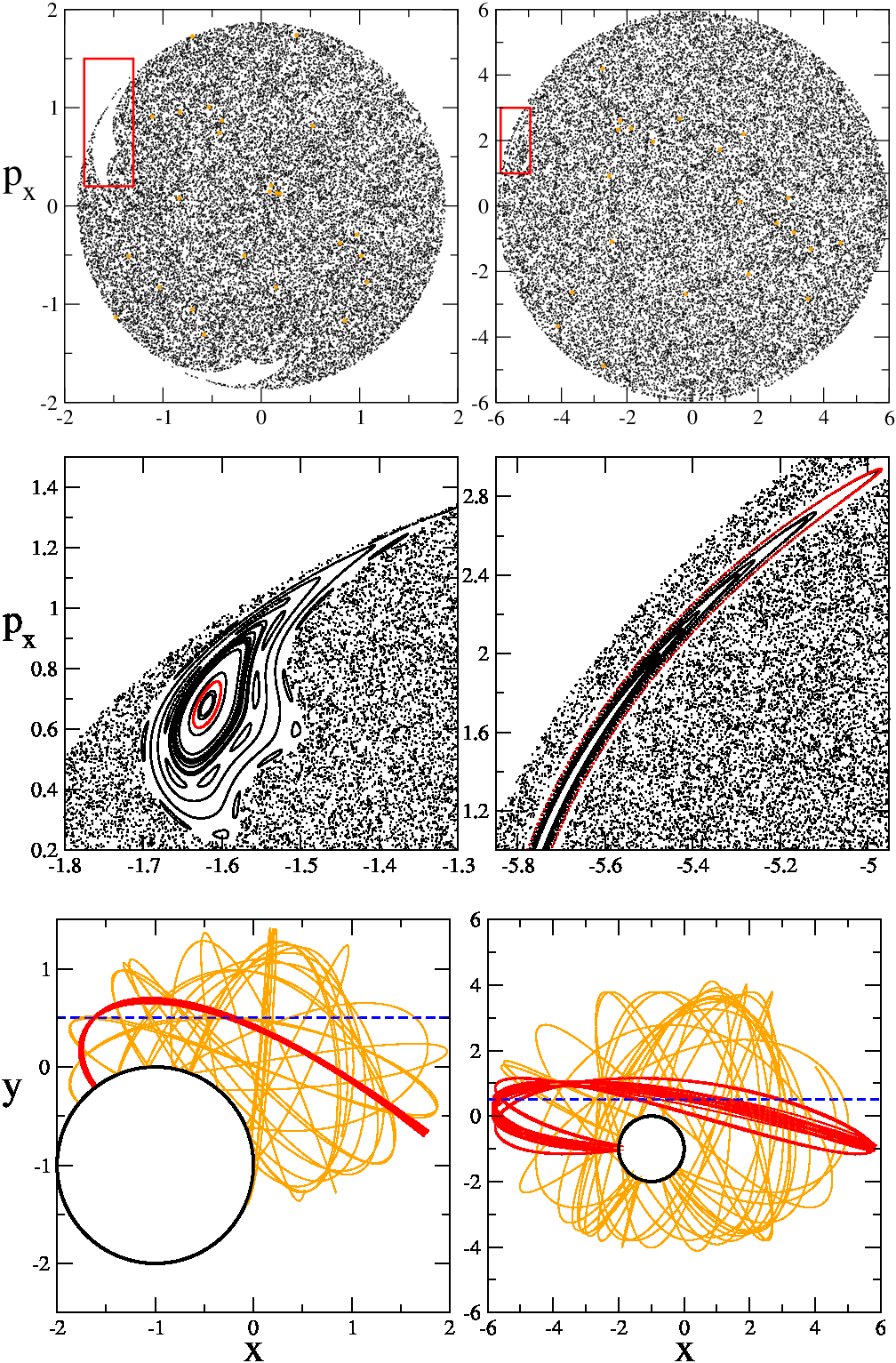}
 \caption{(Color on-line) 
Poincar\'e sections $(x,p_x)$ at $y=0.5, p_y >0$. 
Left and right columns correspond to energy
$E=2$ and $E=18$ respectively. Top panels show the entire 
Poincar\'e sections 
(10 chaotic orbit up to time $t \leq 10^4$)
and middle panels show zoom
marked in top panels 
(adding 15 trajectories in the integrable region to 
time $t \leq 10^4$), one invariant curve of each panel is 
highlighted with red (gray) color inside stability island 
at middle panels.
Bottom panels show dynamics in $(x,y)$ plane
with stable orbits from middle panels
(same red color) and chaotic orbits (orange color also 
shown in top panels) up to times $t = 10^2$; 
dashed horizontal lines 
mark $y=0.5$ used for the Poincar\'e sections.}
\label{fig6}
\end{figure}

Another promising possibility can be an experimental realization of
a harmonic Sinai billiard, or Sinai oscillator. 
An example of such a billiard is described by a 
classical Hamiltonian
$H = ({p_x}^2  + x^2)/2 + ({p_y}^2  + 2 y^2)/2$ with a rigid disk 
of radius $r_d=1$ located at $x=y=-1$ (thus the ratio of 
frequencies in $x$ and $y$ is irrational). The harmonic potential
can be realized by optical traps while the repulsive rigid disk
can be created inside by an additional laser beam
with such a frequency detuning that it acts as a repulsive
potential for cold atoms. Examples of 
typical trajectories in such a billiard are shown in Fig.~\ref{fig6}. 
In the same figure 
we also show the Poincar\'e sections 
constructed in $(x,p_x)$ plane at $y=0.5, \; p_y>0$ 
with energies $E=2$ and $E=18$, when the size of oscillations of atom
is larger than $r_d$. In this regime almost all phase space
is chaotic (the domains of integrable dynamics are very small). 
Thus we think that the GPE in such a harmonic
Sinai billiard will show all the effects of dynamical thermalization
discussed above for a more convential case of the Bunimovich stadium.
We expect that such a system can be more simple for
experimental investigations. Also in such a billiard a coherent state
of the harmonic potential can be created experimentally
and can be used as an initial state with energy being close to the 
energies of linear eigenmodes of such a billiard.
We expect that experimental investigations of the GPE in a harmonic Sinai
billiard (or in the Bunimovich stadium)
will allow to understand the fundamental aspects
of dynamical thermalization.

We thank D.Gu\'ery-Odelin for useful discussions of cold atom physics.

\end{document}